\documentclass[twocolumn,prb,superscriptaddress]{revtex4}

\usepackage{graphicx}
\usepackage{dcolumn}
\usepackage{bm}
\usepackage{amssymb, amsmath}
\usepackage{color}
\usepackage{slashed} 

\begin{document}
	\title{Numerical extrapolation method for complex conductivity of disordered metals }

	\author{S. Kern}
	\affiliation{Department of Experimental
		Physics, Comenius University, SK-84248 Bratislava, Slovakia}
	
	\author{P. Neilinger}
	\affiliation{Department of Experimental
		Physics, Comenius University, SK-84248 Bratislava, Slovakia}
	\affiliation{Institute of Physics, Slovak Academy of Sciences,
		D\'{u}bravsk\'{a} cesta, Bratislava, Slovakia}
	
	\author{D. Manca}
	\affiliation{Department of Experimental
		Physics, Comenius University, SK-84248 Bratislava, Slovakia}

	\author{J. Gregu\v{s}}
	\affiliation{Department of Experimental
		Physics, Comenius University, SK-84248 Bratislava, Slovakia}

	\author{S. Volkov}
	\affiliation{Department of Experimental
		Physics, Comenius University, SK-84248 Bratislava, Slovakia}

	\author{M.~Grajcar}
	\affiliation{Department of Experimental
		Physics, Comenius University, SK-84248 Bratislava, Slovakia}
	\affiliation{Institute of Physics, Slovak Academy of Sciences,
		D\'{u}bravsk\'{a} cesta, Bratislava, Slovakia}

\begin{abstract}
Recently, quantum corrections to optical conductivity of disordered
metals up to the UV region were observed. Although this increase of
conductivity with frequency, also called anti-Drude behaviour, should
disappear at the electron collision frequency, such transition has
never been observed, or described theoretically. Thus, the knowledge
of optical conductivity in a wide frequency range is of great interest.
It is well known that the extrapolation of complex conductivity is
ill-posed - a solution of the analytic continuation problem is not
unique for data with finite accuracy. However, we show that assuming
physically appropriate properties of the searched function
$\sigma(\omega)$, such as: symmetry, smoothness, and asymptotic solution
for low and high frequencies, one can significantly restrict the set
of solutions. We present a simple numerical method utilizing the
radial basis function approximation and simulated annealing, which
reasonably extrapolates the optical conductivity from visible frequency
range down to the far infrared and up to the ultraviolet region. The method was
compared with two widely used analytic continuation techniques and resulting
extrapolation obtained on MoC and NbN thin films was checked by transmission
measurement across a wide frequency range.
\end{abstract}
\maketitle

\section{Introduction}
\label{ch:intro}
Optical properties of thin films, especially superconducting metals,
are currently of great interest. Knowledge of the electric response is
essential for the rapidly expanding field of superconducting devices
e.g. photon detectors,\cite{day2003broadband} as well as for purely
academic purposes, due to the presence of phenomena like weak localization.
Therefore, this class of materials attracts much investigation.
\cite{banerjee2018optical,semenov2009optical,neilinger2019observation}
Measurement of optical properties in a broad frequency range is often
demanding and the experimental window of devices is naturally limited.
However, as it is known, the knowledge of the response function on a finite
interval is sufficient to extend the domain because this function must
be analytical, i.e. satisfying the Kramers-Kronig relations (KK).
Unfortunately, this problem, known as analytic continuation (AC), is ill-posed -
its solution is unstable and not uniquely determined when a small
error is introduced into an input data set.\cite{trefethen2020quantifying}
In order to solve such a problem, regularization should be carried out by making use of additional
information.

For example, one can presume that the studied system can be described by a certain
physical model.\cite{kuzmenko2005kramers,tanner2015use} In the case of metals, the complex conductivity is usually analysed in
terms of the classical Drude-Lorentz model. This model allows to fit the
conductivity of clean metals well, but it fails to describe the conductivity
of disordered metals, where quantum corrections play a significant role\cite{Altshuler1985}
and can persist up to the ultraviolet range.\cite{neilinger2019observation}
The discussion on the applicability of this model is in the Appendix \ref{ch:modind}.
In contrast, there are more general, rather mathematical approaches to the AC,
e.g. the Pad\'{e} approximation and the singular value decomposition (SVD).
However, they exhibit instability due to noise in the data, as we demonstrate
in Appendix \ref{ch:PS}, and the regularization by physically motivated constraints is not straightforward.\cite{schott2016analytic,beach2000reliable,creffield1995spectral}

In this paper, we describe a method which allows to add constraints systematically until
a reasonable extrapolation is obtained. As we demonstrate, our method is able to find the expected
solution, presuming physically well justified restrictions on the searched function. Namely, that
the complex conductivity i) cannot oscillate on scales smaller than the
relaxation rate $\Gamma$, ii) cannot diminish at frequencies higher than $5\Gamma$, iii) has a non-negative real part
and iv) the contribution from high frequency ($\gg\Gamma$) transitions can
be  included in the ”infinite frequency” dielectric constant $\epsilon_\infty$ .
This allows a model-independent determination of $\sigma(\omega)$, extending the
experimental window of spectroscopic ellipsometry by making use of DC sheet resistance,
both measured at room temperature. The approach is demonstrated on data measured on
highly disordered metals close to the metal-insulator transition. Our simple illustrative
procedure is compared to the aforementioned techniques, namely to the Drude-Lorentz data fit,
Pad\'{e} approx. and SVD.

\section{Method description}
\label{ch:metdes}
\begin{figure}
\includegraphics[width=8.6cm]{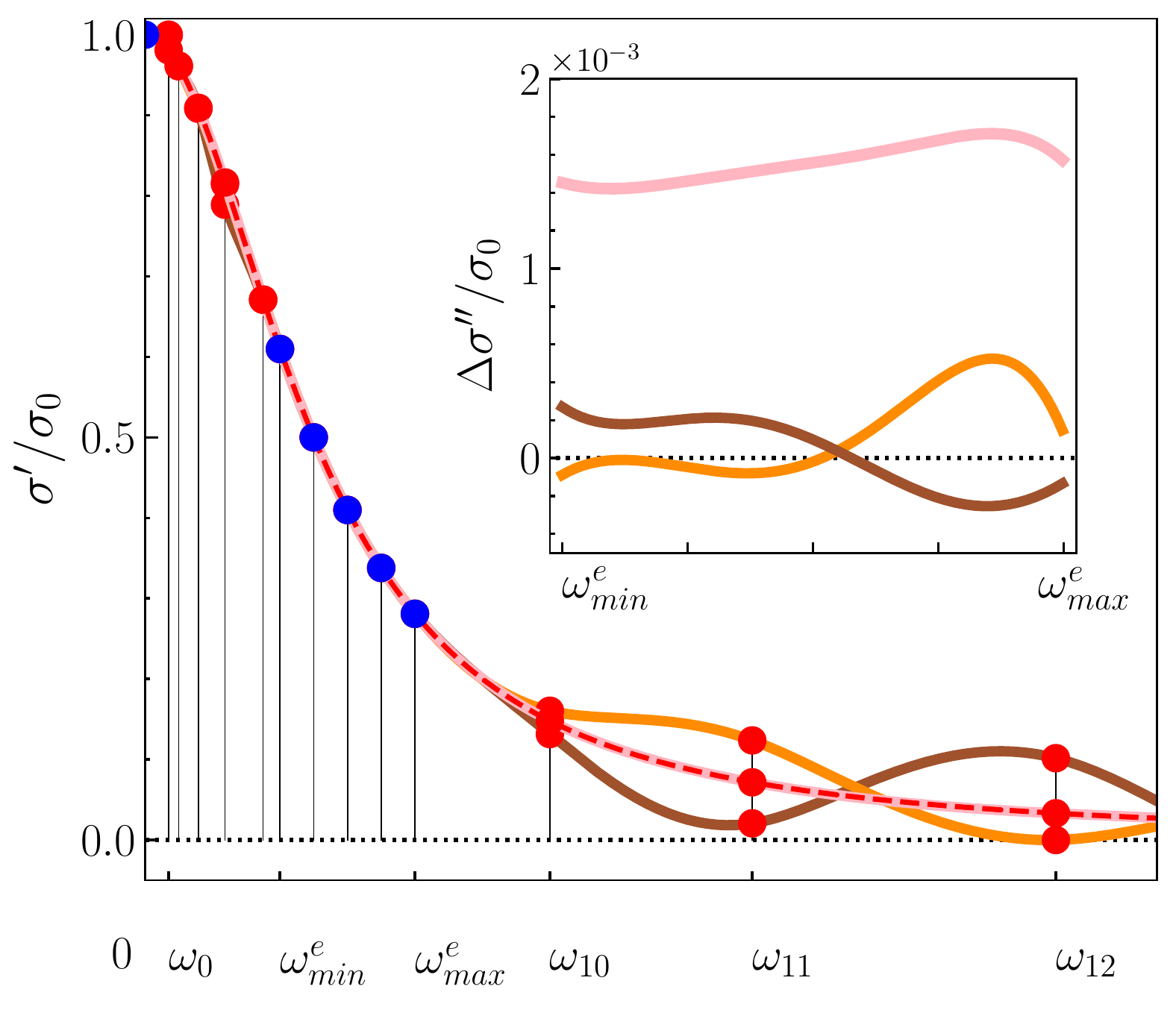}
\caption{Reconstruction of Drude complex conductivity by the described
	method from precisely known values, shown as blue points. The real
	part of Drude conductivity is shown by the red dashed line together with
	three curves, which have considerably different values $y_i$ (red points)
	at higher frequencies. Inset shows the deviation
	$\Delta\sigma^{\prime\prime}/\sigma_0=\mathcal{H}[\sigma^\prime_{\{y_i\}}]/\sigma_0-g^{\prime\prime}$
	of the imaginary part of conductivity calculated from these curves
	by Hilbert transform, from the Drude one. The deviation is surprisingly
	larger for the light pink curve, which has all the red points lying on the real
	part of Drude conductivity, than for the brown and orange curves oscillating
	around it. The values of the functional  (\ref{eq:functional1}) for the light pink
	curve are $\mathcal{F}_0=1.2\times10^{-5}\sigma^2_0$ and for the orange and
	brown curves are $2.0\times10^{-7}\sigma^2_0$. Nevertheless, the average of
	brown and orange curves  is very close to the Drude one.}
\label{fig:Drude}
\end{figure}

Let a set of complex values $\{\sigma^\prime_{e}(\omega_i^e)+i\sigma^{\prime\prime}_{e}(\omega_i^e)\}$
be the complex conductivity at discrete frequencies $\omega_i^{e}$, from
interval $[\omega^{e}_{min},\omega^{e}_{max}]$, as well as at zero frequency,
$\sigma^\prime_e(0)=\sigma_{dc}$, measured at temperature $T_{R}$.
We start with the discretization of the $\omega$ - axis. Taking into account
that the error of analytical continuation of experimental data measured
with finite precision increases exponentially with distance from the
known interval,\cite{trefethen2020quantifying} we use a logarithmic scale.
This also helps to avoid problems with splining square-root and logarithmic
singularities with polynomial functions. We introduce a new discrete variable
$x_i$ with values
\begin{equation}
\label{eq:xvar}
\begin{aligned}
x_i=&0,1,...\quad\quad\\
&...k,\frac{1}{a}\ln(\hbar\omega_0^{e}/k_BT_{R}),\frac{1}{a}\ln(\hbar\omega_1^{e}/k_BT_{R}),...\\
&...,\frac{1}{a}\ln(\hbar\omega_N^{e}/k_BT_{R}),l,l+1,..., n,
\end{aligned}
\end{equation}
where the number of points $k$ below the experimental window is set by a parameter $a=\ln(\hbar\omega_0^{e}/k_BT_{R})/(k+1)$ 
and $l$ is the lowest integer greater than $\ln(\hbar\omega_{max}^{e}/k_BT_{R})$.
The discretization is obtained simply by
\begin{equation}
\label{eq:om_i}
\omega_i=\frac{k_BT_R}{\hbar}e^{a \cdot x_i}. 
\end{equation}
An example of such procedure is shown in  Fig.~\ref{fig:Drude}, where the
exponential spacing of red points $\omega_i$ is interrupted by the experimental
values (blue points) at $\omega^e_i$. Values of the real part of conductivity at the
frequencies $\omega_i$ are denoted by $y_i$, i.e. $y_i=\sigma^{\prime}(\omega_i)$
and they will become the fitting parameters to be optimized. The integer $n$,
determining the number of fitted points, can be estimated as the index of a
reasonably high value of $\omega_n$, where the conductivity can be safely fixed
to zero, since $\sigma(\omega\to\infty)\xrightarrow[]{} 0$, whereas the
contributions from high energy transition, being far enough from the studied
region, are included in parameter $\epsilon_\infty$ introduced later on.
According to Fermi liquid theory, finite temperature can be taken into account
by the transformation
\begin{equation}
\label{eq:trans}
\omega \mapsto \Omega=\sqrt{\omega^2+\gamma(T)^2},
\end{equation}
where $\gamma(T)$ is of the same order of magnitude as $k_BT/\hbar$, often
taken as $\gamma(T)=\pi k_BT/\hbar$ (e.g. Eq.~6.7 in Ref.~\onlinecite{Altshuler1985}).
Equation (\ref{eq:trans}) implies that for $\omega\ll\gamma(T)$, the conductivity
$\sigma^\prime(\Omega)$ is constant and equals to $\sigma_{dc}=\sigma^\prime(\gamma(T))$.
Thus a smooth function $y=f_{\{y_i\}}(x)$ can be  constructed by a cubic spline
using the Radial Basis Function (RBF) method\cite{dyn1983iterative, weber2009b}, with 
two boundary conditions, $y_0=\sigma_{dc}$ measured at room temperature and $y_n=0$. The first condition is
satisfied by fixing the first point of the discretization to frequency 
$\approx\gamma(T_R)$, which is the reason for the presence of the temperature, at which the optical
measurement was conducted, in the discretization (\ref{eq:xvar}). The spline proceeds in
the logarithmic scale, where the distribution of centres $x_i$ given by equation
(\ref{eq:xvar}) is equidistant, which is optimal for RBF method. \cite{iske2000optimal}

From known trial function $f$, the complex conductivity
is calculated as $\sigma_{\{y_i\}}^\prime(\omega)=f_{\{y_i\}}(x(\omega))$ and
$\sigma_{\{y_i\}}^{\prime\prime}(\omega)=\mathcal{H}[\sigma^{\prime}_{\{y_i\}}(\omega)]$.
Here, $\mathcal{H}[\sigma^{\prime}_{\{y_i\}}(\omega)]$
is the Hilbert transform of $\sigma^{\prime}_{\{y_i\}}(\omega)$.
The curve $f_{\{y_i\}}(x)$ is found as the best fit of its Hilbert transform
$\sigma_{\{y_i\}}^{\prime\prime}(\omega)$ to experimental set of points
$\sigma^{\prime\prime}_e(\omega_i^e)$ by least-squares method, minimizing the functional
\begin{equation}
\label{eq:functional1}
\begin{aligned}
\mathcal{F}[\sigma^{\prime}_{\{y_i\}}(\omega)]=\sum_{\omega_j^e} \Big( \sigma^{\prime\prime}_e(\omega_j^e)-\sigma^{\prime\prime}_{\{y_i\}}(\omega_j^e)\Big)^2\\
+\sum_{\omega_j^e}\Big( \sigma^{\prime}_{e}(\omega_j^e)-\sigma^{\prime}_{\{y_i\}}(\omega_j^e)\Big)^2.
\end{aligned}
\end{equation}

This approach leads to the following key observation:
Suppose, for a moment, that the values of $\sigma^\prime_{e}(\omega_i^e)$
and $\sigma^{\prime\prime}_{e}(\omega_i^e)$ were not obtained by measurement,
but were generated instead by restricting the domain of a dimensionless model
function  $g^\prime(\omega)$ and its Hilbert image  $g^{\prime\prime}(\omega)$,
i.e. $\sigma^\prime_{e}(\omega_i^e)=\sigma_0g^\prime(\omega_i^e)$ and 
$\sigma^{\prime\prime}_{e}(\omega_i^e)=\sigma_0g^{\prime\prime}(\omega_i^e)$.
It means that the solution corresponding to the input data $\sigma^\prime_{e}(\omega_i^e)$
and $\sigma^{\prime\prime}_{e}(\omega_i^e)$
is already known (the whole function $g^\prime(\omega)$).
It is clear that the spline of finite number of points $f_{\{y_i\}}(x)$
can not perfectly recover the function, even if $y_i=\sigma_0g^\prime(\omega_i)$.
Slight differences yield small, but non-zero values of the functional (\ref{eq:functional1}),
denoted by $\mathcal{F}_0$. However,by applying the corresponding optimization
method, one can find curves with an even lower value of the functional (\ref{eq:functional1}),
while these curves are significantly different from $g^\prime(\omega)$.

This is interpreted as a practical consequence of the ill-posed nature of the
problem and in order to regularize it, we add the assumption that the searched
function is slowly varying. This can be confirmed by the fact that the relaxation
rate $\Gamma$ in highly disordered metals has a large value (for estimation of
$\Gamma$ see Appendix \ref{ch:estim}) and therefore, their conductivity varies on
large scales.\cite{neilinger2019observation} Utilization of this requirement is based
on the following idea: For the sake of simplicity, let $g^\prime(\omega)$ be the Lorentzian 
\begin{equation}
\label{eq:lor}
g^\prime(\omega)=\frac{1}{1+(\omega/\Gamma)^2}
\end{equation}
shown in Fig.~\ref{fig:Drude} as red dashed line.
The choice of input data $\sigma^\prime_{e}(\omega_i^e)$ and the discretization
of the problem are depicted in Fig.~\ref{fig:Drude} as blue and red dots
respectively. Let us investigate different solutions with the value of functional
(\ref{eq:functional1}) smaller or equal to the $\mathcal{F}_0$. Such solutions
have the following property: if the red point at  $\omega_{11}$
is lifted and the red point at $\omega_{12}$ lowered, or vice versa, the
deviation of the imaginary conductivity from the Drude one, shown in inset
of Fig.~\ref{fig:Drude}, changes only very slightly and can even decrease
the deviation caused by the spline error. Thus, a solution with the least structure can be
obtained by averaging of all generated curves.\cite{gunnarsson2010analytical,schott2016analytic} 
Curves with energy (\ref{eq:functional1}) lower than $\mathcal{F}_0$ can be found, utilizing the simulated annealing
technique by repeated melting of the system. We suggest inverse logarithmic
temperature decay and rather small step in the form of $n$-dimensional
Gaussian random variable, recommended for continuous optimization.\cite{spall2005introduction}
Optimized parameters are forced to be non-negative and the upper bound
is chosen high enough to include the shape of an expected solution but
not too high to make optimization time-consuming.
Points in the measured interval, i.e. values $y_i=\sigma^\prime(\omega_i)$
where $\omega_i \in [\omega^{e}_{min},\omega^{e}_{max}]$
are optimized within intervals $[\sigma^\prime_e(\omega_i^e)-\chi,\sigma^\prime_e(\omega_i^e)+\chi]$,
where $\chi$ is the noise level of the data.

The generated curves $\sigma_{\{y_i\}}^\prime(\omega)$ which satisfy
$\mathcal{F}(\sigma_{\{y_i\}}^\prime(\omega))\leq \mathcal{F}_0$
are used to compute the averaged curve
$\overline{\sigma^\prime}(\omega)=\sigma^{\prime}_{\{\overline{y_i}\}}(\omega)$.
Although the Hilbert transformation is linear, the averaged curve
$\overline{\sigma^\prime}(\omega)$ gives a slightly higher
value of the functional $\mathcal{F}$. Therefore we search a curve varying
similarly to $\overline{\sigma^\prime}(\omega)$, while minimizing the functional
(\ref{eq:functional1}). Such a multi-objective optimization is performed
by linear scalarization,\cite{emmerich2018tutorial} i.e. minimizing functional:
\begin{equation}
\label{eq:functional2}
\tilde{\mathcal{F}}_\lambda[{\scriptstyle\sigma^{\prime}_{\{y_i\}}(\omega)}]=\mathcal{F}[{\scriptstyle\sigma^{\prime}_{\{y_i\}}(\omega)}]+
\lambda\int\displaylimits_{\omega_{0}}^{\omega_{n}}{\scriptstyle \Big( \frac{d^2\sigma^{\prime}(\omega)}{d\omega^2}- \frac{d^2\overline{\sigma^{\prime}}(\omega)}{d\omega^2}\Big)^2}\mathrm{d}\omega,
\end{equation}
where $\omega_0$ and $\omega_{n}$ are the first and the last point
of the spline defined by equation (\ref{eq:om_i}). The parameter $\lambda$
is not \textit{a priori} known and must be carefully chosen for
the particular problem. One can start with a high value of $\lambda$,
such that $\lambda\gg\mathcal{F}/I$, where $I$ is the integral in
(\ref{eq:functional2}), estimated with an arbitrarily chosen curve from
the ensemble. The found solution is very similar to
$\overline{\sigma^\prime}(\omega)$ and with a similar value of the first
error functional (\ref{eq:functional1}). One can lower the value of
$\lambda$ further, until the error (\ref{eq:functional1}) is lower then
$\mathcal{F}_0$. In real data analysis, the precision limit
$\mathcal{F}_0$ is set by the measurement error in such a way, that the
accepted curve lies within the estimated measurement error. 

\begin{figure}[!htbp]
\includegraphics[width=8.6cm]{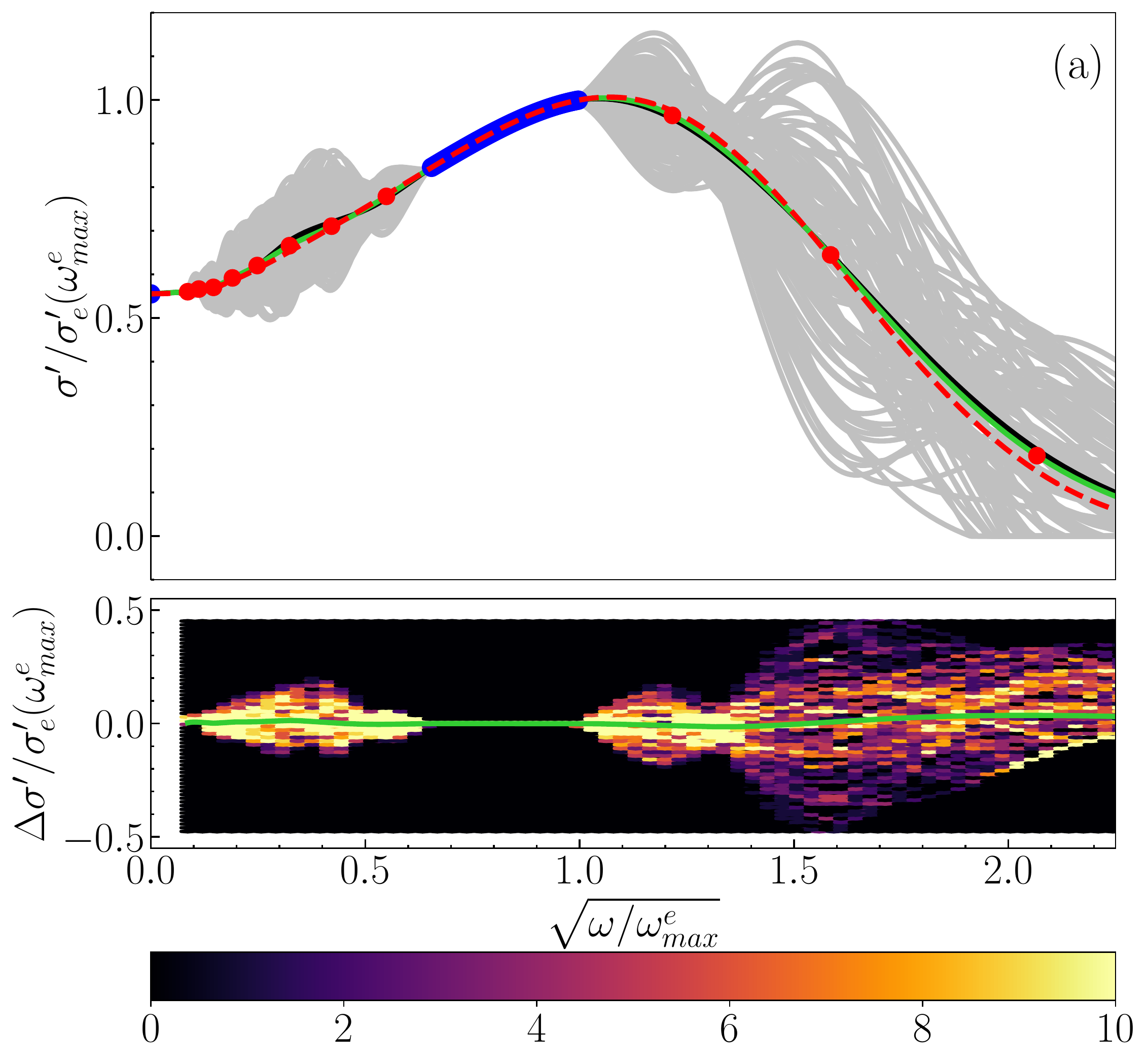}
\includegraphics[width=8.6cm]{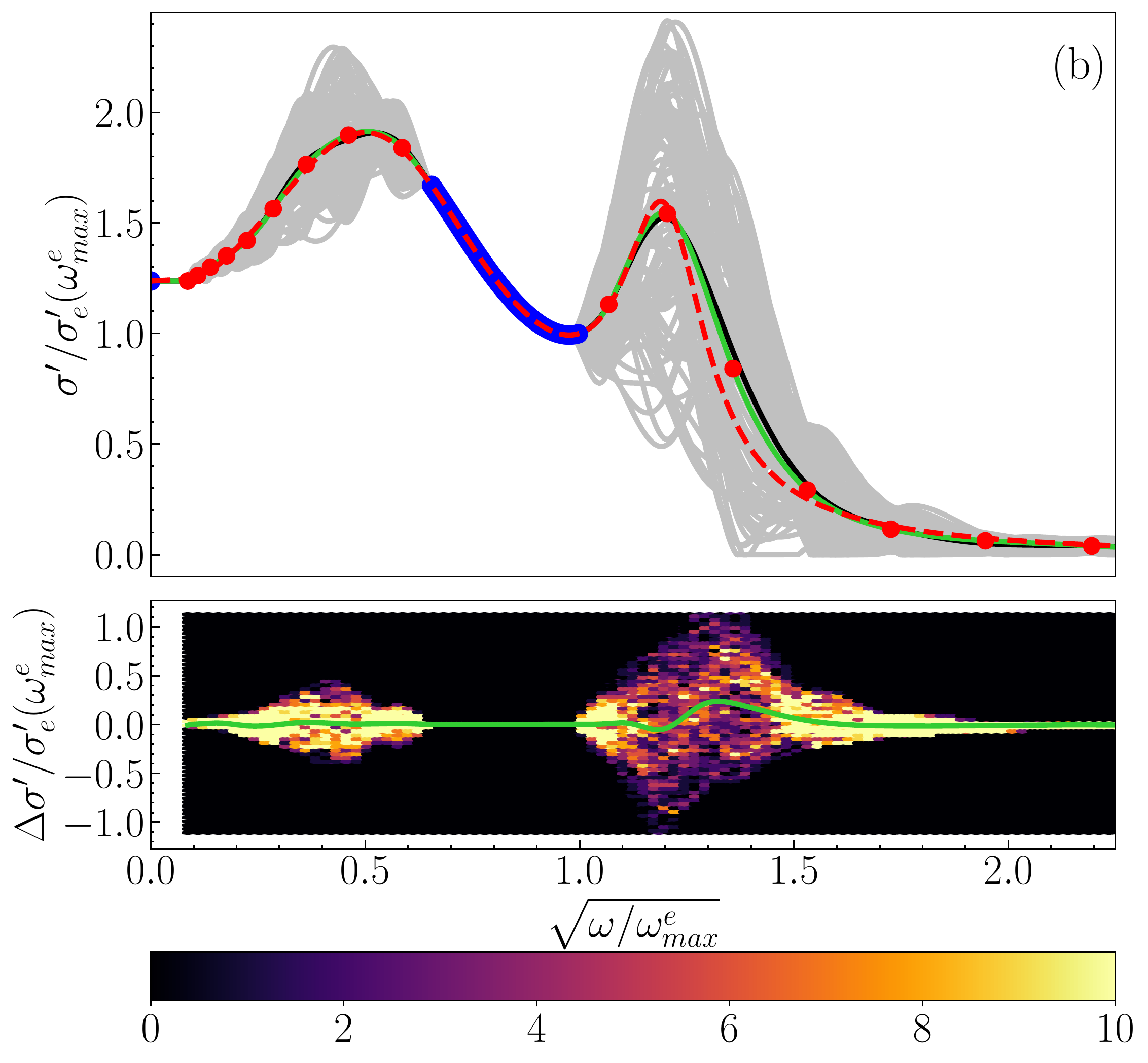}
\caption{(a) Top: Red dashed line is a Gaussian model of optical
	conductivity with square-root quantum corrections (MoC). Blue points
	laying on the curve are input data for the extrapolation. Silver lines
	are curves from the ensemble.  Black line is a normalized averaged
	curve $\overline{\sigma^\prime}(\omega)=\sigma^{\prime}_{\{\overline{y_i}\}}(\omega)$.
	The final curve is green. Bottom: Histogram showing occupancy of
	deviations around the searched function by curves from ensemble and
	deviation of final curve shown as green line. (b) Top: Red dashed line
	is the optical conductivity as a sum of two Lorentzians, with square-root
	quantum corrections (NbN). Blue points are input data, ensemble curves
	are silver, the average curve is black and the final curve is green.
	Bottom: histogram of deviation of curves from ensemble and deviation of
	final green line.} 
\label{fig:2}
\end{figure}
\section{Numerical results}
\label{ch:numres}
The extrapolation range and accuracy of the described procedure, i. e.
finding reasonable complex conductivity curves to extrapolate the experimental
curves  with respect to KK relations, naturally depends on the
degree of complexity of the extrapolated functions. Thus, to demonstrate the
feasibility of this method, we tested the extrapolation on sets of data created
from two functions, which qualitatively describe the real part of the optical
conductivity of MoC and NbN thin films, respectively. The conductivity of
the former is modeled by a simple function motivated by Ref.~\onlinecite{neilinger2019observation},
which contains the observed square-root quantum corrections to the optical
conductivity of these films:
\begin{equation}
\label{eq:moc}
g_{1}(\omega,T)=e^{-\Omega^2/\Gamma_{1}^2}+Q^2(\sqrt{\Omega/\Gamma_1}-1)e^{-4\Omega^2/\Gamma_1^2}.
\end{equation}
Here, $\Omega$ is defined by equation (\ref{eq:trans}), $\Gamma_1$ is the
scattering rate, and the quantumness $Q^2$ characterizes the strength of the
square-root corrections, which are significant up to a certain crossover
frequency, chosen to be half of the scattering rate. This function is shown
as the red dashed line in square-root-scale in top plot of Fig.~\ref{fig:2}(a)
for $Q^2=0.66$. The input data - depicted as blue dots -  were obtained by
sampling the above function (\ref{eq:moc}) at $20$ regularly placed frequencies.
Points of discretization are depicted as red dots. Next, to test the
extrapolation of a slightly more complex function, we study a model function
with two peaks motivated by Refs.~\onlinecite{semenov2009optical,neilinger2019observation} 
\begin{equation}
\label{eq:nbn}
\begin{aligned}
g_{2}(\omega,T)=\frac{1}{1+(\Omega/\Gamma_{2,1})^2}+&Q^2(\sqrt{\Omega/\Gamma_{2,1}}-1)e^{-4\Omega^2/\Gamma_{2,1}^2}\\
+&\frac{r_{2}}{1+((\Omega-\Omega_{2})/\Gamma_{2,2})^2},
\end{aligned}
\end{equation}
where $\Gamma_{2,1}/2$ is the cut-off for quantum corrections and $r_{2}$ is
the peak height ratio. Here, the cut-off frequency was not determined by the $\Gamma_{2,1}$
but by the appropriate scale describing the ultimate decrease of the function to zero, which
we assumed to be $\Omega_2+\Gamma_{2,2}$. As we show in the Appendix \ref{ch:estim}, the estimated
relaxation rate $\Gamma$ for the NbN film corresponds rather to this scale. The searched function (\ref{eq:nbn})
is shown in the top plot of Fig.~\ref{fig:2}(b) as the red dashed line, the input data are blue
dots and the optimized points are red. A few curves from the averaged ensemble
are also shown as silver lines. The average curves are depicted as black lines
and the final conductivity is green. As shown in section (\ref{ch:realdata}),
function (\ref{eq:nbn}) based on Drude-Lorentz model\cite{semenov2009optical}
qualitatively describes the measured conductivity of our NbN films, which
indicates the presence of two peaks in optical conductivity.

The results presented in Fig.~\ref{fig:2} show the usability of the procedure. 
For the $g_1(\omega)$ model, there is good agreement of the final green curve with
the original curve (\ref{eq:moc}) at both higher and lower frequencies. The
deviation $\Delta\sigma^\prime=g_1-\sigma_{y_i}^\prime$ shown in bottom plot
of Fig.~\ref{fig:2}(a) is below $5\%$. Shown is also a histogram generated by
the ensemble of found curves. The conductivity
model  $g_2(\omega)$ is more complex, therefore, the function (\ref{eq:nbn}) is accurately
reconstructed in the vicinity of the extrapolated region, whereas further from this region,
the deviation increases. The deviation of the final curve from the searched function (\ref{eq:nbn}),
shown in bottom plot of Fig.~\ref{fig:2}(b), reaches $20\%$ and variance of the curves is also larger
than for the simpler model (\ref{eq:moc}). This is visible at higher frequencies in
the slightly underestimated second peak's height, whereas at low frequencies,
the function is recovered with lower deviation. In both
cases, we performed both the Pad\'{e} approx. and SVD. The results are presented and discussed in Appendix \ref{ch:PS}.

\section{Extrapolation of experimental data}
\label{ch:realdata}
\begin{figure}
  \includegraphics[width=8.6cm, height=6.4cm]{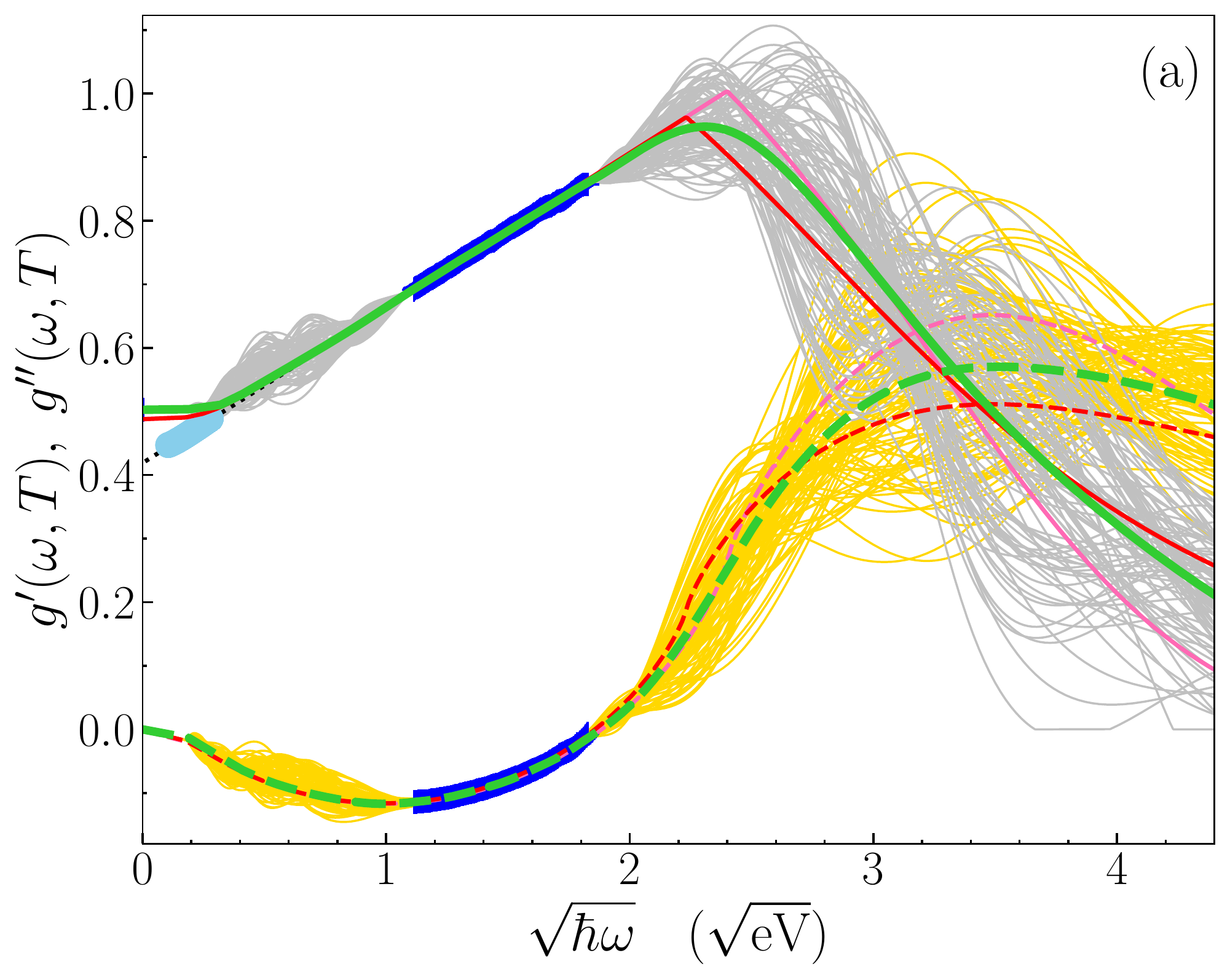} 
  \includegraphics[width=8.6cm, height=6.4cm]{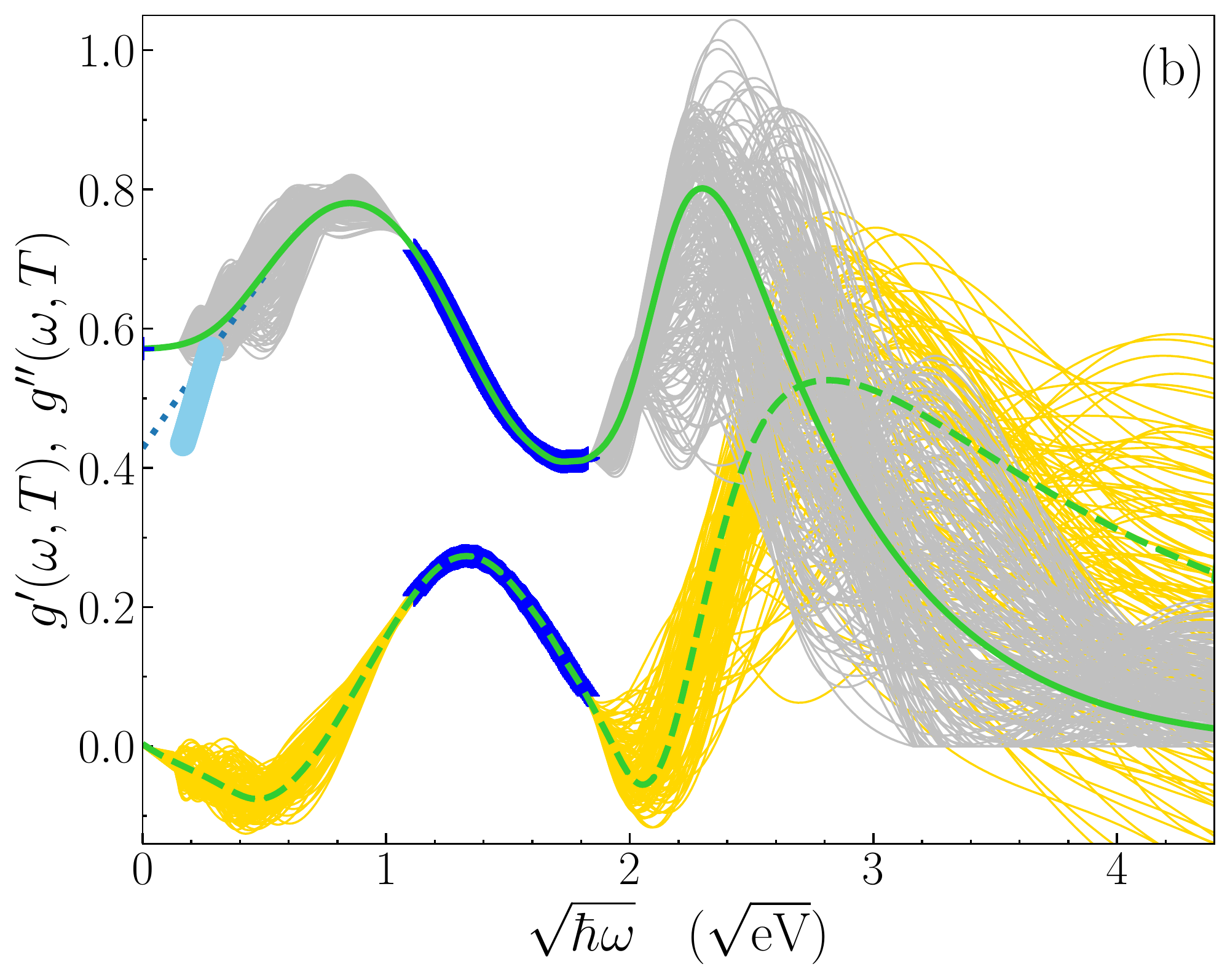}
 \caption{Real (solid green line) and imaginary (dashed green line) parts of the
 	 extrapolation of normalized complex sheet conductance
 	$g=g^\prime+ig^{\prime\prime}=(\sigma^\prime+i\sigma^{\prime\prime})dZ_0$, obtained
 	from spectroscopic ellipsometry (blue data) for MoC (a) and NbN (b) thin films with
 	thicknesses $d=5$~nm and $3.5$~nm, respectively, measured at room temperature.
 	All found complex conductivities with functional (\ref{eq:functional1})
 	$\leq\mathcal{F}_0$ are shown as silver (real part) and gold lines (imaginary part).
 	Since Eq. (\ref{eq:trans})  implies $g^{\prime}(\omega=0,T)=g^{\prime}(\omega=\pi k_BT,0)$,
 	the real part of complex conductivity at zero temperature (dotted line),
 	obtained from extrapolation procedure, can be compared to temperature-dependent
 	DC transport measurement shown as light blue thick lines. Plot (a) also shows two
 	theoretical curves, a Lorentzian (red) and a Gaussian (pink),
 	proposed in Ref.~\onlinecite{neilinger2019observation}.   }
\label{fig:3}
\end{figure}

The procedure was applied to extrapolate the conductivity of disordered thin
films, as shown in Fig.~\ref{fig:3}(a) for MoC and Fig.~\ref{fig:3}(b) for NbN, 
respectively. For the details of sample preparation, see Appendix \ref{ch:sampprep}
and Refs.~\onlinecite{neilinger2019observation, volkov2019superconducting}.
Complex conductivity was determined from spectroscopic ellipsometry, and
DC conductivity was measured by the Van der Pauw method. The contribution of
interband transitions at higher energies must be taken into account in the
imaginary part of the conductivity. Following the KK relations, the contribution
from the bound electrons can be expressed as
$\sigma^{\prime\prime}_{bound}=-\epsilon_0(\epsilon_\infty-1)\omega$, where
$\epsilon_0$ is the permittivity of vacuum and $\epsilon_\infty$
is the bound-electron contribution to the static dielectric constant. For
MoC, the value of $\epsilon_\infty$ was estimated as 1.4 in
Ref.~\onlinecite{neilinger2019observation} and for NbN, we calculated
$\epsilon_\infty=2.6$ utilizing the same procedure, while a similar value was
estimated in Ref.~\onlinecite{semenov2009optical}. Subsequently, the corresponding
contribution to the imaginary part of the conductivity $\sigma^{\prime\prime}_{bound}$
was subtracted from measured data. Since the relaxation rate $\Gamma$ can not be easily
used as a fitting parameter, we estimated the value from the sheet resistance
measured on our thin films and the density of carriers. The details on both estimations
of $\epsilon_\infty$ and $\Gamma$ are given in the Appendix \ref{ch:estim}.
For MoC conductivity, the extrapolation fits between two theoretical curves, a Lorentzian (red) and a Gaussian (pink),
proposed in Ref.~\onlinecite{neilinger2019observation}. Taking $\gamma(T)$ in
the form $\pi k_B T / \hbar$ (same as in Ref.~\onlinecite{neilinger2019observation}),
the inverse transformation given by the equation (\ref{eq:trans}) allows to
compare the extrapolation (dotted lines) with the terahertz-frequency real part
of conductivity (light blue thick lines), obtained from the temperature dependent
DC  transport measurement. The agreement for MoC is very good, especially taking
into account the large distance between optical frequencies (500 THz) and the
frequency ($\pi kT/h\approx 50$~THz) relevant for transport measurements. The
agreement for NbN is less satisfying. However, even for these more complicated
conductivity spectra, the extrapolation can match the transport measurements reasonably.

\begin{figure}
 \includegraphics[width=8.6cm, height=6.4cm]{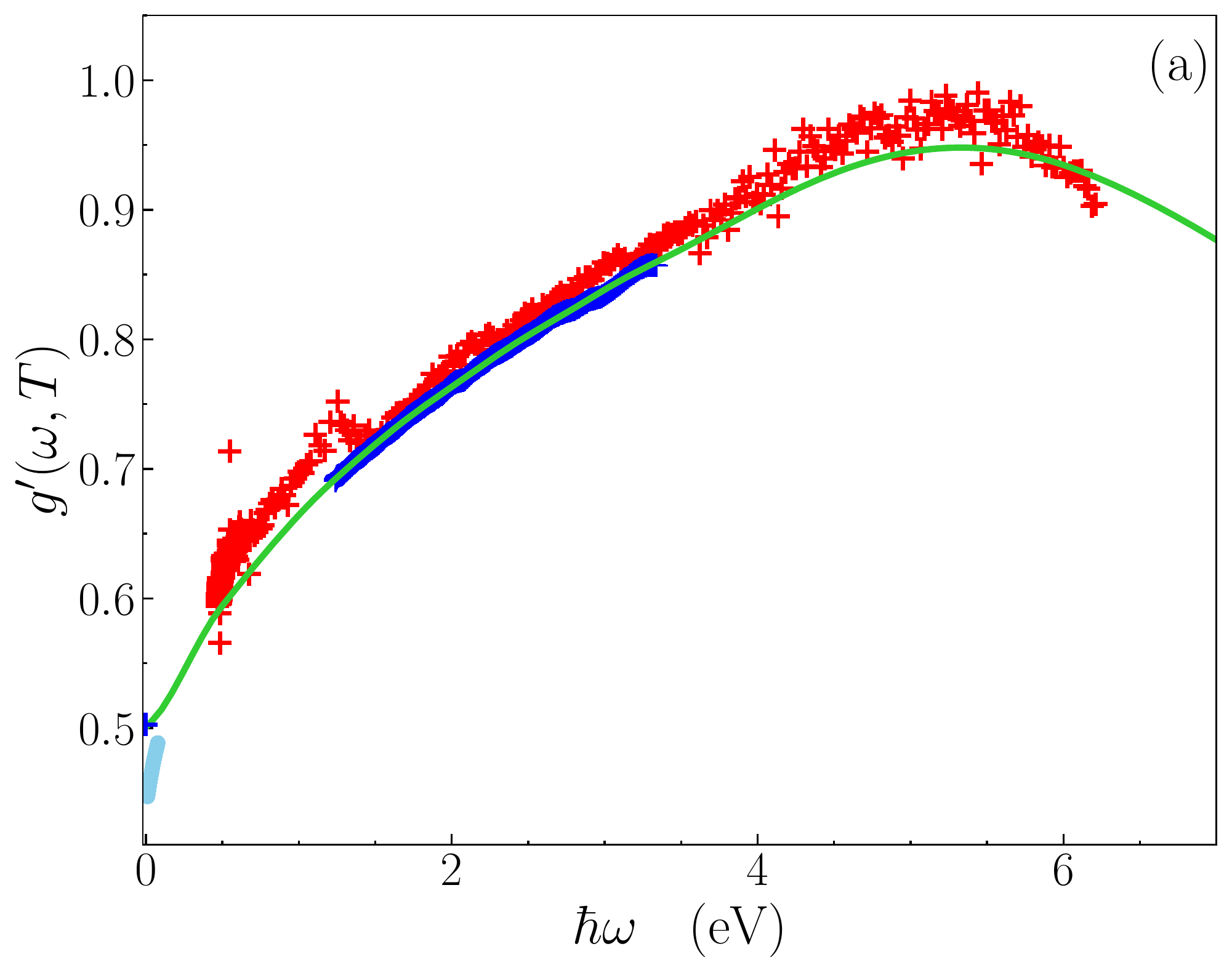} 
 \includegraphics[width=8.6cm, height=6.4cm]{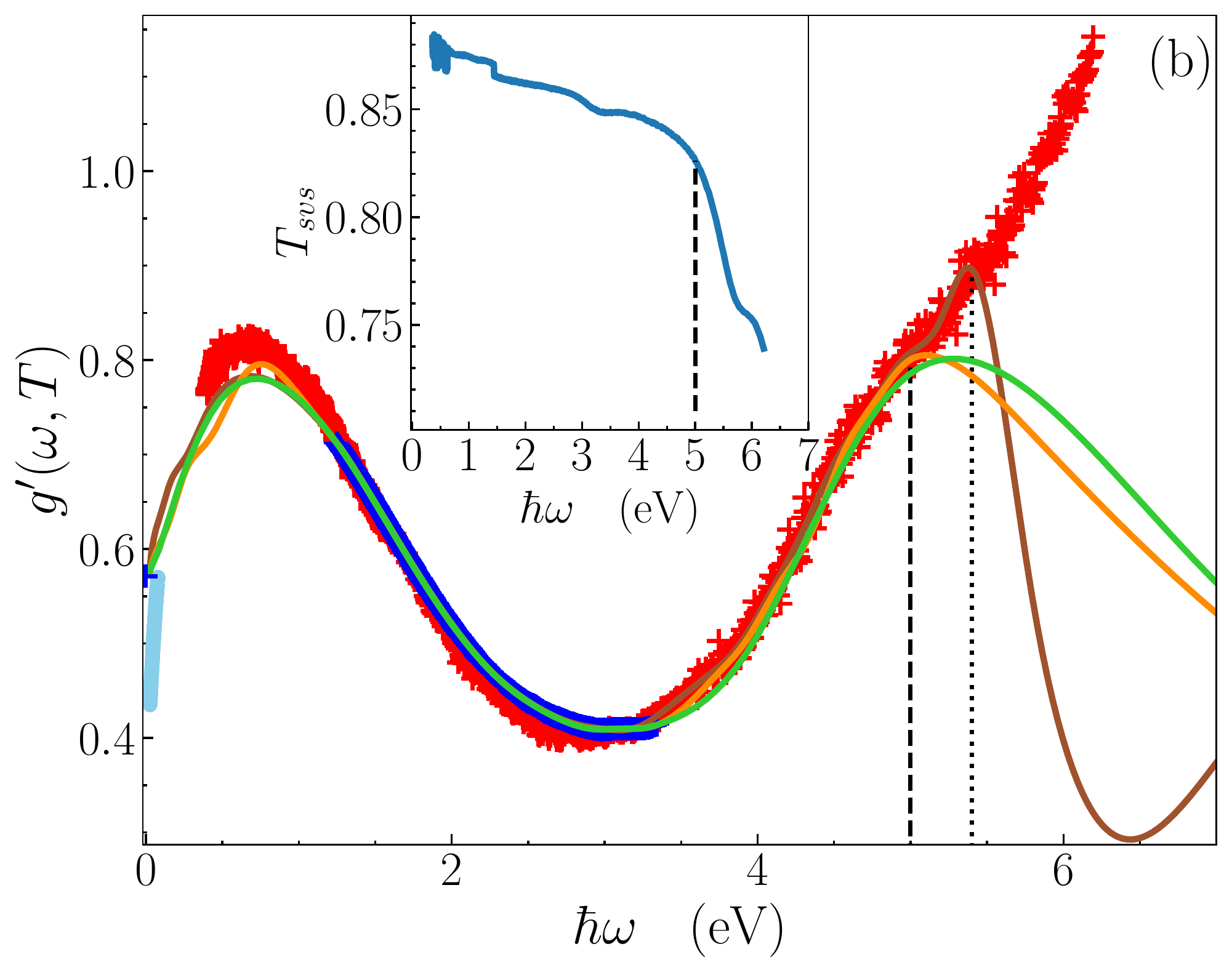}
\caption{Green curves are extrapolations of normalized sheet conductance for MoC (a) and
	NbN (b) obtained from spectroscopic ellipsometry data (blue points) (the same as in
	Fig.~\ref{fig:3})  and the experimental values obtained from transport (light blue)
	and  transmission (red) measurements. In plot (b) brown and orange curves are extrapolations
	of data obtained from transmission measurement with upper cut-off 5.4~eV and 5.0~eV,
	respectively. The value 5.4~eV of the energy cut-off (dotted line) is the largest,
	where the extrapolation can be done. The highest value of the cut-off energy, where
	the influence of the substrate can be neglected, is 5.0~eV. The rapid change in the
	brown curve above 5.0~eV is probably due to the influence of the substrate, whose
	transmission changes rapidly above this value, as shown in the inset. Data are shown in linear scale. }
\label{fig:4}
\end{figure}

In order to verify the extrapolation procedure experimentally, the real part of
conductivity, determined from the extrapolation of ellipsometric data, was compared
to the conductivity calculated from the transmission of MoC and NbN films (see Appendix \ref{ch:optprop}),
directly measured in the much wider frequency range accessible by our spectrometer,
as shown in Fig.~\ref{fig:4}. The good agreement of our extrapolation with the data from transmission measurement for
MoC is the encouraging result. The mismatch in the NbN's transmission above 5.0~eV
could be caused by interdiffusion between sapphire and NbN. The transmission of
sapphire is nearly frequency independent between 0.3-5.0~eV  (inset in Fig.~\ref{fig:4} (b)),
but above 5.0~eV, the transmission strongly depends on impurities in the
sapphire.\cite{dobrovinskaya2009sapphire} To verify the reliability of transmission
data above 5.0~eV, we used them to compute the real part of conductivity, taking the
imaginary part from ellipsometry  for the same sample, and used both as the
basis for extrapolating a complex conductivity curve, plotting its real part (Figure 4).
Such extrapolation failed if the entire transmission dataset (6.2~eV) was used,
indicating that such conductivity violates the Kramers-Kronig relations. By the
bisection method, we found the cut-off, for which the transmission data still process
into valid complex conductivity, to be 5.4~eV (dotted line, brown); if a data set with
cut-off at 5.0~eV was used, the real part of the generated conductivity (orange) matches
that produced from ellipsometry alone (green).

\section{Conclusion}
\label{ch:con}

A numerical procedure of extrapolating the complex conductivity of disordered metals
based on the Kramers-Kronig relations was introduced. In general, this task has many
solutions with standard methods finding a single particular one. The proposed method finds a single
solution by averaging many of them which is reasonable, if the requirement of slow
variation of their conductivity on energy scale $\lesssim\Gamma$ is met. A slow variation of $\sigma(\omega)$
is reasonably satisfied in highly disordered metals, where experimental data do not
indicate rapid oscillations in conductivity and the described method offers a robust and
reliable extrapolation procedure of the measured optical conductivity. The range and
precision of the extrapolation depend on the complexity of the function. For the simple
single-peak model of conductivity, which is the case for MoC thin films, even the DC
transport measurements can be reconstructed from the optical measurements, despite the
presence of strong quantum corrections. For the more complex, double-peaked conductivity
model, which is the case of NbN films, the reliability range of the extrapolation is
decreased. However, the extrapolation still reasonably matches the DC measurements and
predicts the second peak.

\subsection{Acknowledgments}

This work was supported by the Slovak Research and Development Agency under
the contract APVV-16-0372,  APVV-18-0358 and by the QuantERA grant SiUCs, by SAS-MTVS.

\newpage

\appendix
\section{Optical properties calculation}
\label{ch:optprop}
Both the ellipsometry data and the extrapolated conductivity were
used to obtain the optical properties of our thin films. The real
and imaginary part of the complex refractive index are defined in
Ref.~\onlinecite{hecht2002optics} as
\begin{equation}
\label{eq:reffr_ind}
\begin{aligned}
&n(\omega)=\sqrt{\frac{1}{2}(\Re\{\epsilon_R(\omega)\}+\lvert\epsilon_R(\omega)\rvert)} \\ 
&k(\omega)=\textrm{sgn}(\omega)\sqrt{\frac{1}{2}(-\Re\{\epsilon_R(\omega)\}+\lvert\epsilon_R(\omega)\rvert)},
\end{aligned}
\end{equation} 
where $\epsilon_R$ is the (complex) relative permittivity given by
\begin{equation}
\label{eq:con_eps}
\epsilon_R(\omega)=1+i\sigma(\omega)/\omega\epsilon_0.
\end{equation}
The transmission $T_{vfs}$ of a thin film with the complex refractive
index $\tilde{n}(\omega)=n(\omega)+ik(\omega)$, placed upon
a semi-infinite substrate with index $\tilde{n}_s=n_s$, i.e. vacuum-film-substrate,
whose imaginary part is neglected is \cite{li2018optical}
\begin{equation}
\label{eq:tvsf}
T_{vfs}=\Big\lvert\frac{2}{(1+\tilde{n}_s)\textrm{cos}\phi-i(\tilde{n}_s+\tilde{n}^2)\phi_0\frac{\textrm{sin}\phi}{\phi}}e^{-i\phi_0}\Big\rvert^2.
\end{equation}
Here $\phi=\tilde{n}\frac{\omega}{c}d$ and $\phi_0=\frac{\omega}{c}d$ are
phase shifts in the film and in vacuum of thicknesses $d$, respectively.
The transmission $T_{vfsv}$ through a system with finite substrate thickness is
\begin{equation}
\label{eq:tvfsv}
T_{vfsv}=\frac{T_{vfs}T_{sv}}{1-R_{vfs}R_{sv}},
\end{equation}
where $T_{sv}=4n_s/(1+n_s)^2$ is the transmission of the substrate-vacuum
interface, and the reflection $R$ is obtained from the corresponding transmission
simply by $1-T$. Next, the transmission of the vacuum-substrate-vacuum system is
\begin{equation}
\label{eq:tvsv}
T_{vsv}=\frac{T_{sv}^2}{1-R_{sv}^2}.
\end{equation}
Finally, the equations (\ref{eq:tvfsv}) and (\ref{eq:tvsv}) are used to express
the transmission of the system vacuum-film-substrate-vacuum normalized
to the transmission of the substrate itself as
\begin{equation}
\label{eq:tn}
\frac{T_{vfsv}}{T_{vsv}}=\frac{T_{vfs}}{T_{sv}}\frac{1-R_{vs}^2}{1-R_{vfs}R_{sv}},
\end{equation}
which was also a measured quantity. The right-hand side of equation (\ref{eq:tn})
is a function of complex conductivity marked as $T_n(\sigma^\prime,\sigma^{\prime\prime})$
and can be utilized to calculate one of its parts, real or imaginary, from another
part and from normalized transmission. Instead of inverting equations
(\ref{eq:reffr_ind}) - (\ref{eq:tn}) one can simply minimize the difference between
$T_n(\sigma^\prime,\sigma^{\prime\prime})$ and measured $T_{vfsv}/T_{vsv}$, namely
\begin{equation}
\label{eq:sigma_from_T}
\sigma^\prime=\arg\min\Big((T_n(\sigma^\prime,\sigma^{\prime\prime})-T_{vfsv}/T_{vsv})^2\Big),
\end{equation}
which is, in fact, plotted in Fig.~(4) as red symbols.

\section{Sample preparation}
\label{ch:sampprep}
The MoC thin film was prepared by means of reactive magnetron
sputtering from Mo target in argon-acetylene atmosphere (both gases
used of purity 5.0) on c-cut sapphire wafer heated to 200 degrees
Celsius.  The flow rates of argon and acetylene were set by flow meters. 
During deposition, the magnetron current was held
constant at 200 mA, corresponding to a deposition rate $\approx 11$~nm/min.
The deposition time, and thus the thickness, was regulated by means of a
programmable shutter control interface with a precision of 1~s. The
chamber was evacuated to approximately $5\times 10^{-5}$~Pa. For 
details on the preparation of the MoC films and their characterization
see Ref.~\onlinecite{trgala2014superconducting}. The sheet resistance 
of the studied MoC sample with a nominal thickness $d=5$~nm was
$R_\Box= 720~\Omega$.

The thin NbN film was prepared by pulsed laser ablation from an Nb-target
(purity $99.99~\%$) in atmosphere of $\textrm{N}_2$ with added $1~\%$ $H_{2}$
(purity of the gas mixture is $5.0$). The NbN thin film was deposited on 
c-cut sapphire wafer, heated to the $600\ ^\circ\textrm{C}$. The used laser
fluency of the KrF laser of $4.9~\textrm{Jcm}^{–2}$ corresponded to the deposition
rate of $2.4~ \textrm{nm}/\textrm{min}$. The vacuum chamber was evacuated to
$2\cdot10^{–6}~\textrm{Pa}$ before deposition. For more details of the
preparation of NbN film and its growing features see.\cite{volkov2019superconducting}
The sheet resistance of our $3.5$~nm thick NbN films was $R_\Box= 655~\Omega$.

\section{Standard methods of extrapolation}
\label{ch:modind}
In the following section, the standard methods of the complex conductivity extrapolation
are applied to the MoC and NbN noisy experimental data and compared to our method. 
\subsection{Fitting to model function}
\label{ch:DL}
\begin{figure}
	\includegraphics[width=8.6cm]{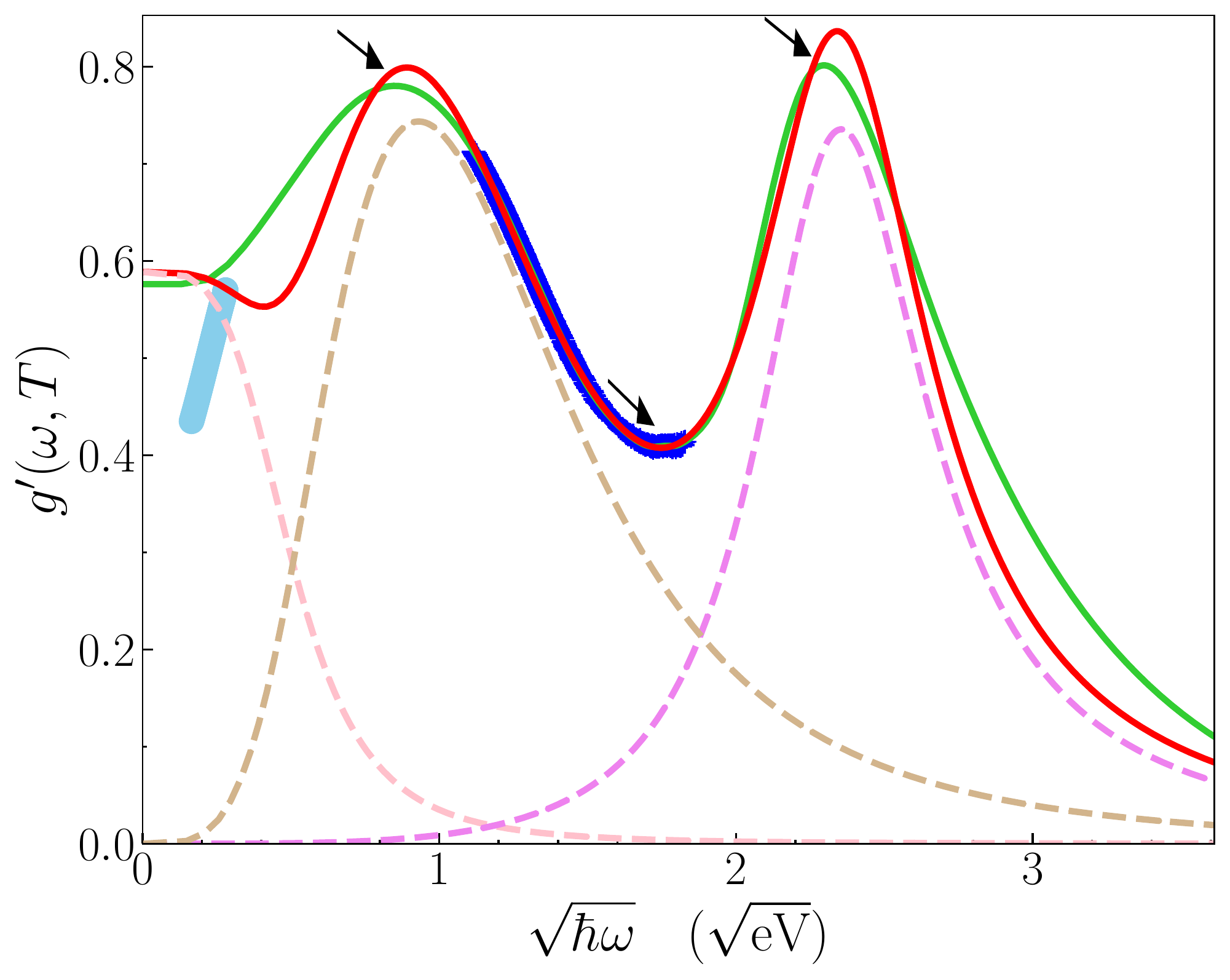}
	\caption{Fit of the normalized sheet conductance (blue dots), by the
		Drude-Lorentz model (red line). The fit does not match the increase of the
		conductivity indicated by the temperature-dependent DC transport measurement
		(light blue thick line) in contrast to the prolongation (green line), see Fig.~4 (b)
		in the main text. The pink, brown and violet dashed lines show the contributions
		from the Drude and the two Lorentz terms in (\ref{eq:DL}), respectively.} 
	\label{fig:fit}
\end{figure}

The presumed model determines a direct physical interpretation of the parameters of the fit, which have a direct physical interpretation, are determined 
by . The ill-posed nature of the analytic continuation enables one to fit the finite
interval data with reasonable accuracy by a variety of models.  
Therefore, the final solution should be verified by an independent measurements.   
For example, the conductivity of metallic thin films is commonly 
analysed by the Drude-Lorentz model.\cite{Ketterson2016}, as presented  by Semenov et al. in
Ref.~\onlinecite{semenov2009optical} for NbN thin film:
\begin{equation}
\label{eq:DL}
\epsilon_r=\epsilon_{\infty}-\frac{\omega_p^2}{\omega(\omega+i\Gamma)}+\sum_{n=1}^2\frac{\Omega_{Sn}^2}{\Omega_{0n}^2-\omega^2-i\Omega_{Dn}}
\end{equation}
Here, $\epsilon_r$ is the relative permittivity from which the conductivity is
obtained by equation \eqref{eq:con_eps}. This function does not take into consideration the
presence of quantum corrections to the optical conductivity, which are clearly
present in highly disordered NbN films. These corrections can strongly affects
the optical conductivity of these films, depending on the degree of disorder. 
The contribution of quantum corrections to optical conductivity should be present
up to UV frequencies, as assumed in theory and recently experimentally demonstrated
in Ref.~\onlinecite{neilinger2019observation}. Still, such contributions are usually
neglected above infra-red frequencies and the complex conductivity is commonly 
analysed by the Drude-Lorentz model. This leads to either unrealistic physical parameters
or to the conclusion that the complex conductivity cannot be fitted by the standard
model.\cite{mArcher2017} The fit of our NbN data by the Drude-Lorentz model \eqref{eq:DL}
is shown in Fig.~\ref{fig:fit}. The first Lorentz peak, $i=1$ in (\ref{eq:DL}), which should
corresponds to a band transition, is misused to fit the conductivity suppressed by
quantum corrections at lower frequencies and the normal regime at higher frequencies.
As a consequence, the Drude contribution is suppressed to create a crossover to the low
frequency conductivity and the fit provides an extremely small value of the relaxation
rate $\hbar\Gamma\approx 0.26~eV$ in comparison with rough estimation $\hbar\Gamma\approx10~eV$
presented in next section. In fact, all quantities the listed in Tab I are unphysical, as was explicitly
pointed out in Ref.~\onlinecite{semenov2009optical}, too.  Moreover, fitting data by an
inappropriate physical model can lead to the appearance of additional features that do
not have to be real. On the contrary, our "unbiased"  method  has predictive strength and
can uncover phenomena present at frequencies outside the measured interval, just assuming
that the optical conductivity cannot rapidly oscillate on a scale determined by the relaxation
rate $\Gamma$, whereas rapid changes are not excluded. This is shown in Fig.~\ref{fig:fit}
by arrows which mark frequencies, where optical conductivity changes its monotonicity. Due to
quantum corrections, the change can be very sharp, e.g. thermally smeared square-root or
logarithmic singularity at zero frequency. 

\begin{table}
	\centering
	\label{tab:1}
	\begin{tabular}{|l|l|l|l|l|l|l|l|} 
		\hline
		$\omega_p\hbar$ & $\Gamma\hbar$ & $\hbar\Omega_{S1}$ & $\hbar\Omega_{01}$ & $\hbar\Omega_{D1}$ & $\hbar\Omega_{S2}$ & $\hbar\Omega_{02}$ & $\hbar\Omega_{D2}$  \\ 
		\hline
		\multicolumn{8}{|c|}{(eV)}                                                                                                                                     \\ 
		\hline
		2.88            & 0.25          & 9.42        & 0.86           & 2.12          & 11.71          & 5.54          & 3.31                                         \\
		\hline
	\end{tabular}
	\caption{Parameters given by the fit of the real and imaginary part
		of the complex conductivity obtained from spectroscopic ellipsometry
		on our NbN sample by model (\ref{eq:DL}). We assumed $\epsilon_\infty=2.6$, coming from
		calculation as mentioned in the main text.}
\end{table}

\subsection{Pad\'{e} approximation and SVD}
\label{ch:PS}

\begin{figure}
	\includegraphics[width=8.6cm]{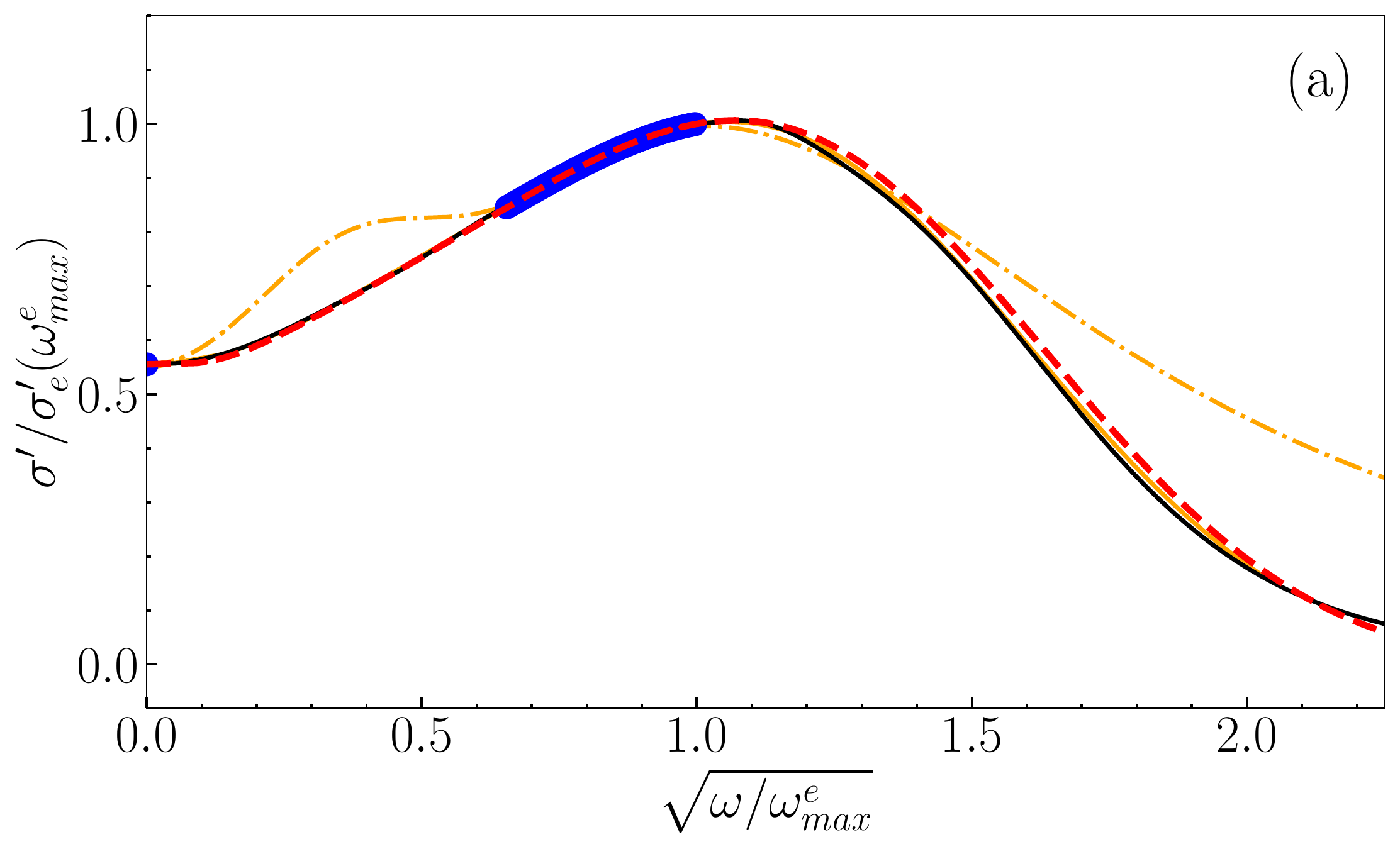}
	\includegraphics[width=8.6cm]{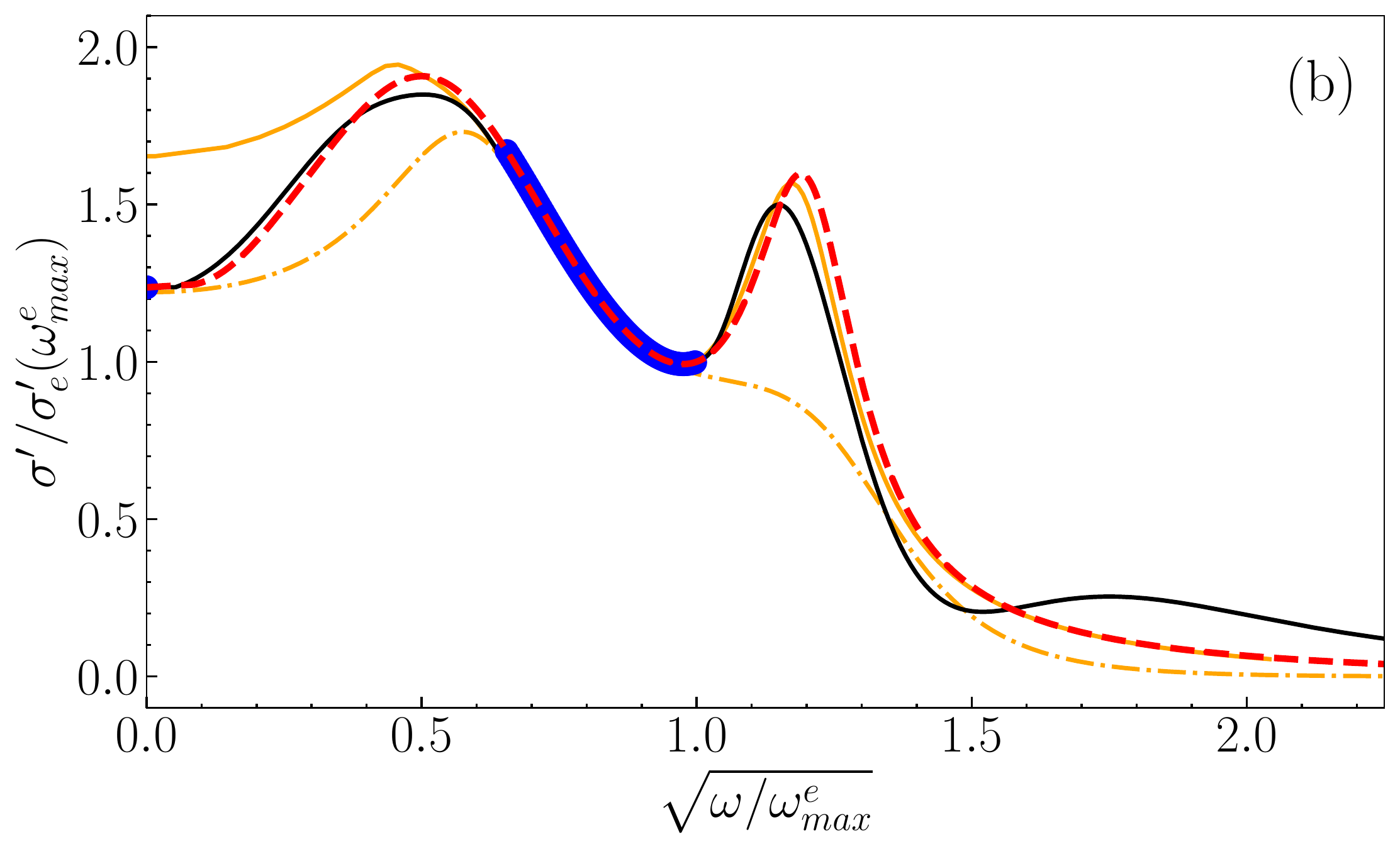}
	\caption{ Application of the Pad\'{e} and SVD prolongation method on the model data (blue dots,
		same as in Fig.~\ref{fig:2}). The red dashed line in panel (a) and (b) corresponds to the model function (7) and (8), respectively.
		Black line is real part of the solution obtained by SVD, orange dash-dotted line is Pad\'{e} approx. with both boundary conditions included and orange solid line is Pad\'{e} approx. without them.} 
	\label{fig:num_std}
\end{figure}

\begin{figure}
	\includegraphics[width=8.6cm]{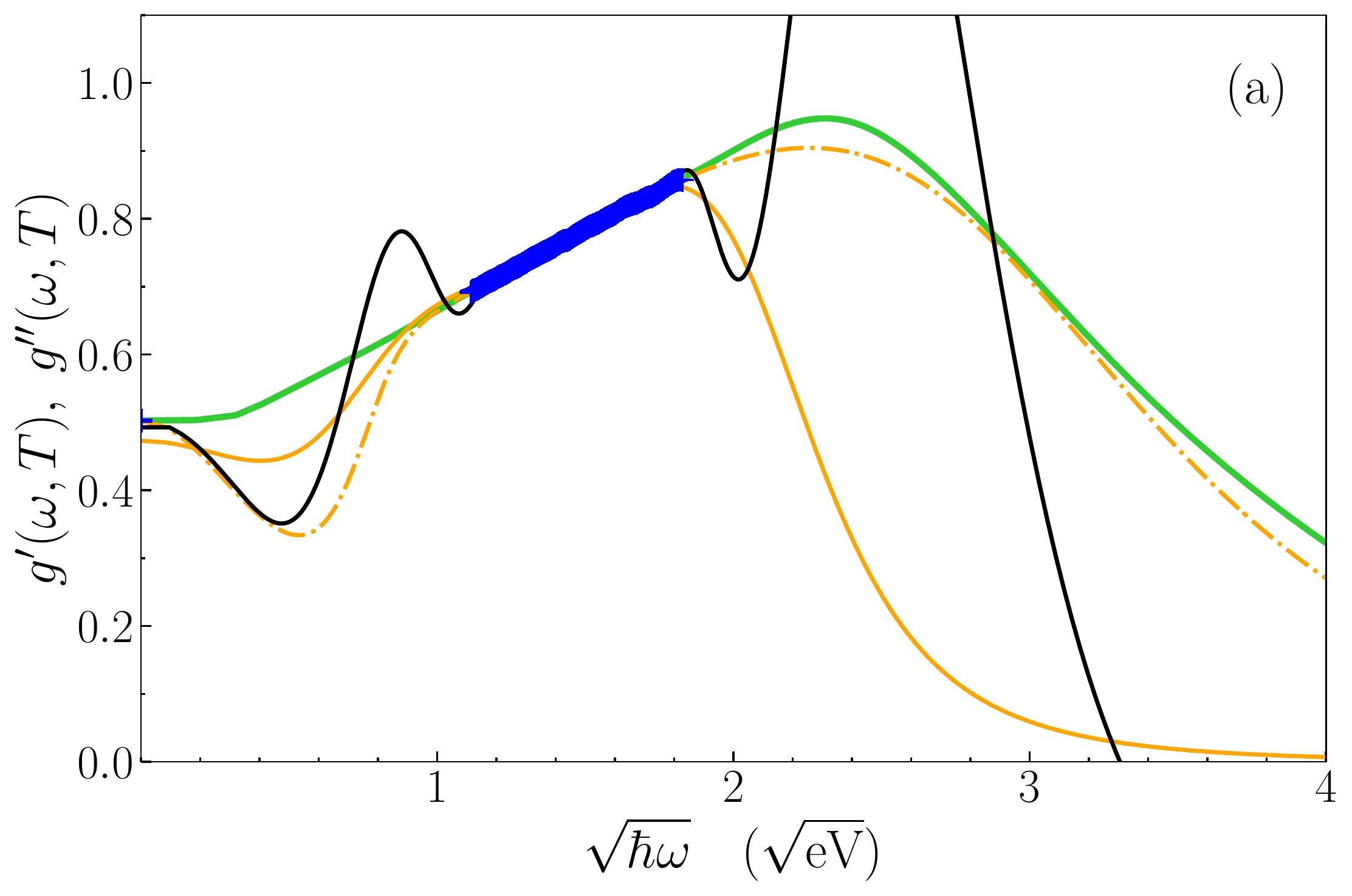}
	\includegraphics[width=8.6cm]{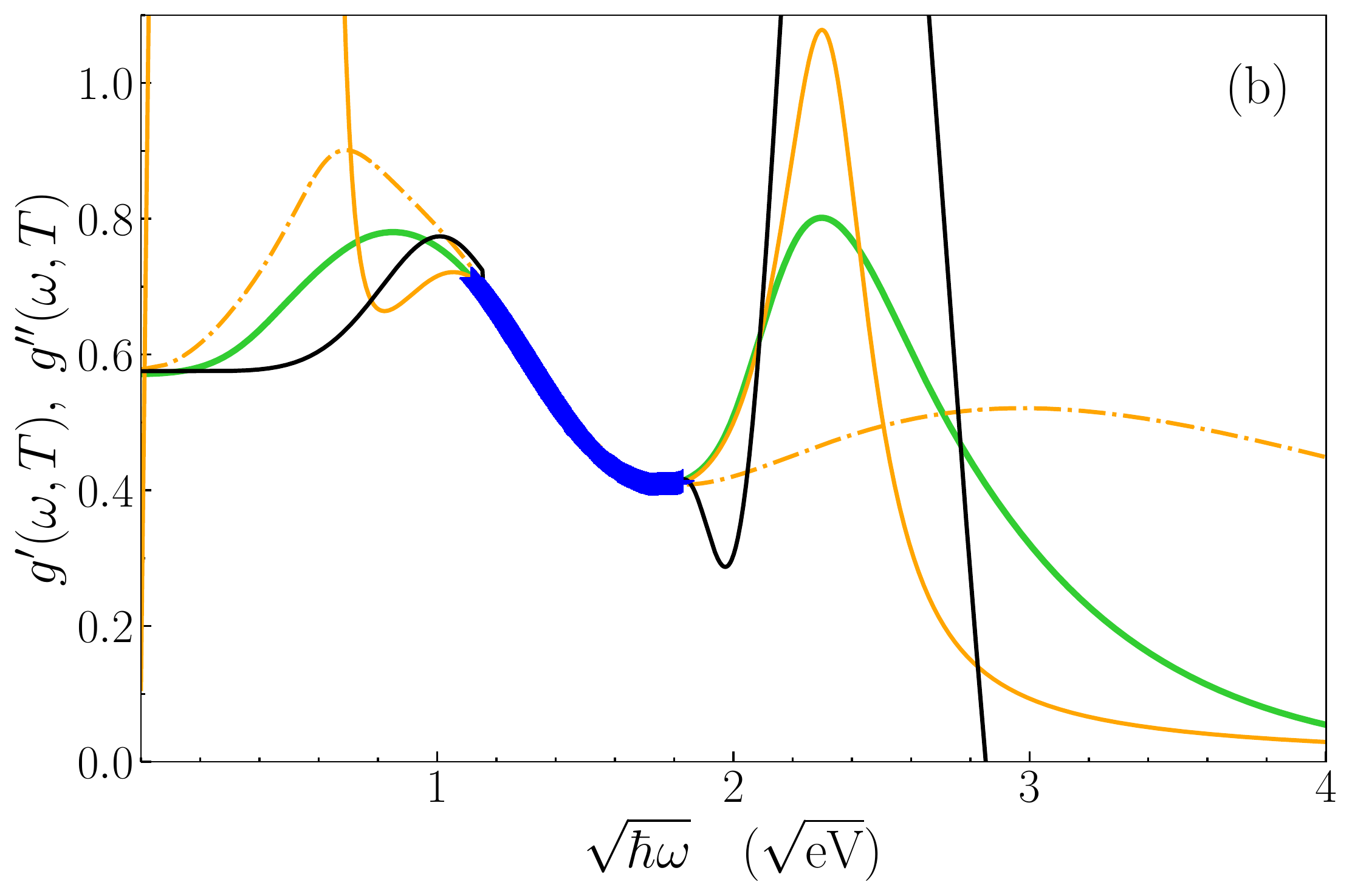}
	\caption{Comparison of the real parts of normalized sheet conductance obtained by our method
		(green line) extrapolating ellipsometric (blue) data to the solutions
		obtained by the Pad\'{e} approx. (orange solid), constrained Pad\'{e} (orange dash-dotted) and SVD (black).} 
	\label{fig:exp_std}
\end{figure}

In the Pad\'{e} approximation, the data are fitted by the ratio of two general
complex polynomials, $\sigma(\omega)=P(\omega)/Q(\omega)$ with orders $N$ and
$M$, respectively. The first boundary condition is satisfied by including the 
point at zero frequency and the asymptotic behaviour at higher frequencies is
ensured by setting $M=N+1$.\cite{schott2016analytic} Results of such constrained Pad\'{e} approx.
strongly depends on the orders $N$ and $M$, which were chosen to avoid reaching
negative, unphysical values. This approach offered poor precision of extrapolation
(dash-dotted line in Fig.~\ref{fig:num_std}) even for the best fit, which was obtained for the orders
$N=5$ and $M=6$ for model (\ref{eq:moc}) and $N=4$ and $M=5$ for model (\ref{eq:nbn}). Surprisingly, fitting
data in the interval $[\omega^{e}_{min},\omega^{e}_{max}]$ only (solid orange curve), without condition on the orders, offered the best result. For
the single peak model (\ref{eq:moc}) the maximum deviation reached was $~1\%$. For the double peak
function (\ref{eq:nbn}), the result was similar but only in higher frequencies. Non-negativity
was again ensured by the orders $N$ and $M$, which were optimized to reach the best
match to the value at zero frequency. For the unconstrained Pad\'{e}, they were both
equal to $6$. The Pad\'{e} approx. which worked reliably for the noiseless model
data, was applied to the MoC and NbN noisy experimental data as well (Fig.~\ref{fig:exp_std}). Although the
noise was carefully filtered out, the unconstrained  Pad\'{e} approx. (solid orange curve) rapidly
varies and diverges for the NbN data. The orders of polynomials were chosen to reach the best
match to the static conductivity, being $N=5$, $M=6$ and $N=8$, $M=9$ for MoC and
NbN, respectively. The constrained Pad\'{e} (dash-dotted line) gave a solution which was more stable,
but poorly fitted the data, especially at the edges of the experimental window.

In SVD, the behaviour of the searched function is prescribed at both boundaries.
Following Ref.~\onlinecite{dienstfrey2001analytic}, the problem of finding
a function satisfying the KK relations, while fitting the input data, is turned
into solving a linear system by expanding the known and searched functions into a
proper series. Regularization is done by truncating the series and the resulting
system is solved by the SVD, which means that the solution is one with
the least squares error. For the model function (\ref{eq:moc}),  a result similar to Pad\'{e}
was achieved by SVD, with the rank of the decomposed matrix being equal to 7. The
rank was chosen to minimize the rapid oscillations of the resulting curve. For
the double peak model (\ref{eq:nbn}), SVD recovers the searched function at low frequencies
very well, but at high frequencies gives additional structure. To obtain this result the
matrix rank was equal to 8. The results obtained from SVD are shown in Fig.~\ref{fig:num_std} as black lines.

When applied to the experimental data, the SVD solution oscillates, its real part
reaching negative values, which can not be forbidden in the SVD approach.
The rank of the matrix of which we calculated the pseudo-inverse by SVD was
again chosen to minimize the oscillations, even though they are still present in
the final curves (black lines in Fig.~\ref{fig:exp_std}). This rank was 7 and 6 for MoC and NbN, respectively.

We believe, that the local variability of
the curves generated by our method allows us to suppress the effects of the
noise and averaging inside the measured interval provides additional smoothing of
data, but with respect to KK.

\section{Estimation of dielectric constant $\epsilon_\infty$ and relaxation rate $\Gamma$}
\label{ch:estim}
Extrapolating the complex conductivity as described requires two crucial parameters,
electron relaxation rate $\Gamma$ and an "infinity frequency" dielectric constant $\epsilon_\infty$.
Our estimation of $\epsilon_\infty$ closely follows calculation presented in Ref.~\onlinecite{neilinger2019observation}:
\begin{equation}
\label{eq:eps_inf}
\epsilon_\infty\approx 1+\sum \frac{\Omega_j^2}{\omega_j^2} .
\end{equation}
Here, approximate equality means that the high frequency permittivity consists of resonant
contributions of transitions between bands modeled by atomic-like levels and conduction
band at the Fermi energy. Estimation of weights of the oscillator was based on the $f-$sum
rule as $\Omega_j^2=\frac{Z_jn_{at}e^2}{m\epsilon_0}$, where $Z_j$ is the number of electrons
at the atomic level $j$ and $n_{at}$ is the density of the particular atom type. The excitation
energies of the atomic levels $\hbar\omega_j$ were taken from Ref.~\onlinecite{gwyn2013electron}
and corresponding densities were calculated from volumes of elementary cells, which can be
found in Ref.~\onlinecite{osti_1201157,osti_1201682}. The calculation for MoC yields
$\epsilon_{\infty}=1.38$.\cite{neilinger2019observation} Relevant levels of niobium are 4p (6 electrons, 31~eV),
4s (2 electrons. 56~eV) and 3d (10 electrons, 203~eV) and levels of nitrogen are 2s
(2electrons, 20~eV) and 1s (2 electrons, 410~eV) which yields  $\epsilon_{\infty}=2.58$ for NbN.

Our rough estimation of the relaxation rate $\Gamma$ is based on the Drude formula for static conductivity
\begin{equation}
\label{eq:drude_static}
\Gamma=\frac{e^2n}{m\sigma_{0}}.
\end{equation}
Here, the density of carriers $n\approx 4.2\times10^{23}~cm^{-3}$ is taken from the Hall
measurement  in Ref.~\onlinecite{destraz2017superconducting}. The quantum corrections to
the static conductivity can be included via the quantumness $Q^2\approx1/{(k_Fl)^2}$,
which relates the classical value $\sigma_0$ to the measured one as
$\sigma_{dc}=\sigma_0(1-Q^2)$.\cite{neilinger2019observation} For a reasonable quantumness\cite{neilinger2019observation}
$Q^2=0.7$ we obtained $\hbar\Gamma\approx16~eV$ and $\approx10~eV$ for MoC and NbN thin film,
respectively. Here we should emphasize, that our method requires a rough estimation of the
relaxation time (order of magnitude) only. For example, if $\Gamma$ is overestimated, the method
sets more high frequency points close to zero conductivity.  On the other hand, if $\Gamma$ is considerably
underestimated, the simulated annealing finds no curve.  In each case, one can properly adjust $\Gamma$
to obtain reasonable curves found by the simulated annealing.


\end{document}